\documentclass[a4paper,12pt,prb,amsmath,amssymb,floatfix,preprint,showpacs]{revtex4}

\newcommand{\eg}{{\it e.g.\ }}
\newcommand{\ie}{{\it i.e.\ }}

\usepackage{longtable}
\usepackage{lipsum}
\usepackage[usenames,dvipsnames]{color}
\usepackage[pdftex]{graphicx} 
\DeclareGraphicsExtensions{.pdf}
\usepackage{color}
\usepackage{array}
\newcolumntype{L}[1]{>{\raggedright\let\newline\\\arraybackslash\hspace{0pt}}m{#1}}
\newcolumntype{C}[1]{>{\centering\let\newline\\\arraybackslash\hspace{0pt}}m{#1}}
\newcolumntype{R}[1]{>{\raggedleft\let\newline\\\arraybackslash\hspace{0pt}}m{#1}}
\begin{document}
\author{H. C. Herper$^{1}$, T. Ahmed$^{2}$,  J. M. Wills$^{3}$, I. Di Marco$^{1}$, T. Bj{\"o}rkman$^{4}$,
D. Iu\c{s}an$^{1}$, A. V. Balatsky$^{2,5}$ and O. Eriksson$^{1,6}$}
\address{$^{1}$Department of Physics and Astronomy, Uppsala University, Box 516, 751 20 Uppsala, Sweden \\
$^{2}$ Institute for Materials Science, Los Alamos National Laboratory, Los Alamos, NM  87545, USA\\
$^{3}$Theoretical Division, Los Alamos National Laboratory, Los Alamos, NM 87545, USA\\
$^4$\AA bo Akademi, Department of Natural Sciences, FIN-20500 Turku, Finland\\
$^5$Nordita, Stockholm, Sweden\\
$^6$School of Natural Science and Technology, \"Orebro University, SE-70182 \"Orebro, Sweden\\
}
\pacs{71.15Mb, 71,20Dg, 72.15Rn, 74.70Xa}
\begin{abstract}
Recent progress in materials informatics has opened up the possibility of 
a new approach to accessing properties of materials in which one assays the aggregate properties of a
large set of materials within the same class in addition to a detailed investigation
of each compound in that class. Here we present the first large scale investigation
of electronic properties and correlated magnetism in Ce-based compounds
accompanied by a systematic study of the electronic structure and 4f-hybridization
function of a large body of Ce compounds.  
We systematically study the electronic structure and 4f-hybridization function of a 
large body of Ce compounds with the goal of elucidating the nature of the 4f states 
and their interrelation with the measured Kondo energy in these compounds.
The hybridization function has been analyzed for more than 350 data sets (being part of the IMS data base) of cubic Ce 
compounds using electronic structure theory that relies on a full-potential approach.
We demonstrate that the strength of the hybridization function, 
evaluated in this way, allows us to draw precise conclusions about the degree of localization of the 4f states in these compounds.
The theoretical results are entirely consistent with all experimental information, relevant to the degree of 4f localization
for all investigated materials.
Furthermore, a more detailed analysis of the electronic structure and the hybridization function
allows us to make precise statements about Kondo correlations in these systems.
The calculated hybridization functions, together with the corresponding density of states, reproduce the expected exponential behavior of the observed Kondo temperatures and prove a consistent trend in real materials. 
This trend allows us to predict which systems may be correctly identified as Kondo systems.
A strong anti-correlation between the size of the hybridization function and the volume of the systems has been observed.
The {\em information entropy} for this set of systems is about 0.42.
Our approach demonstrates the predictive power of materials informatics when a large number of materials is used to establish significant trends.
This predictive power can be used to design new materials with desired properties. 
The applicability of this approach for other correlated electron systems is discussed.
\end{abstract}
\title{Combining electronic structure and many-body theory with large data-bases: a method for predicting the nature of 4f states in Ce compounds\\}
\maketitle
\section{Introduction}
The field of materials informatics is undergoing rapid change, driven by our ability to analyze data sets for a large collection of compounds.
Examples of this approach include the Materials Project\cite{matproj}, 
the Materials Properties Database from NIST\cite{matprop}, 
the Organic Materials Database,\cite{orgmat} 
and the Materials Web\cite{matweb}. 
Here we present the first effort to apply the informatics approach to Ce-based compounds 
with the goal of extracting systematic correlations between electronic structure and properties such as Kondo scaling and volume in {\it f}-electron materials. 
In performing this analysis we use the recently developed f-electron database. \cite{IMS}
We focus on cerium, the most abundant rare earth element, that in various compounds has been used in many applications. 
CeO$_2$, for example, is used for catalysis in combustion engines~\cite{Kaspar:99} and as an abrasive in glass and lens manufacturing.\cite{Brooks:92}
More recently Ce has been used in laser technology, \eg in crystal lasers (Li-Sr-Al-F-Ce) which are used to detect air pollution.\cite{Coutts:04}
Ce is also used for steel hardening processes and in Al coatings.\cite{Keis:63, Kreshchanovskii:63,Harvey:13}
Depending on the application, Ce-based compounds with different materials properties are needed, and since the electronic structure
is often decisive in determining these properties, understanding of the degree of localization/delocalization of the 4f shell is crucial.
In fact, it is important to understand the degree of localization/delocalization of the f-shell, in general, for
lanthanide- and actinide-compounds.
Improvements in materials properties are expected from this knowledge,
\eg in rare-earth-lean permanent magnets where the contribution from 4f states to the magnetic anisotropy energy is generally
large, and determined by the nature of the 4f states.

Ce compounds have been investigated in great detail experimentally as well as theoretically 
and numerous applications have been found.\cite{Reinhardt:12} 
However, identifying new materials for a special purpose or with certain materials properties is a bit like searching for a needle in a haystack.
We have performed extensive high-throughput calculations in which we have linked the electronic properties of the materials or to by precise the hybridization energy
to a tuneable, materials specific parameter such as the volume.
As we demonstrate, this allows us to identify, among all known cubic Ce compounds, the correct level of itineracy of the 4f shell.
This predictive power is likely to carry over also to other f-electron systems.

We note that the search for criteria with which to
gauge the itineracy of the 4f shell has been underway for some time.
The so called Hill plot has often been used for this purpose. Hill identified a correlation between magnetism, superconductivity, 
and the interatomic distance in f-based systems (Ce and U).
A beautiful relationship was found such that
smaller atomic distances were found in compounds that had superconductivity,
whereas materials with larger distances displayed magnetism.\cite{Hill:70}
To be precise, Hill observed that superconductivity did not occur for compounds with a Ce-Ce distance larger than 3.4 {\AA}, and that magnetism did not occur for materials with a Ce-Ce distance smaller than 3.4 {\AA}.
For U compounds, the critical distance (the {\it Hill-limit}), was 3.5 {\AA}.
Although the correlation observed by Hill was excellent, some  exceptions have been found.
CeRh$_3$B$_2$ and URh$_3$,
for example, both have large distance between Ce (U) atoms,
but the expected magnetism is absent in both compounds.
This has motivated the search for
other measures that involve interatomic distances in Ce based compounds, in order to predict
properties (see \eg Ref.\,\onlinecite{Sereni:95, deLong:91}). 
Here we extend this  approach with the use of a large data set from our database.

Simply stated, the degree of localization of the {\it f}\thinspace-electrons
is determined by the competition between energy gain due to band formation,
and energy cost due to
the fact that when itinerant electrons move through the lattice,
they sometimes occupy the same lattice site, with an increased Coulomb repulsion, the Hubbard U.
This competition in interactions may be quantified via the Hubbard model,
which, in the dynamical mean field approximation~\cite{Metzner:89, Georges:92},
can be solved  in terms of the Anderson impurity problem.
In this approach, the Hubbard U enters naturally, and the band formation may be translated into a hybridization function,
\ie the hybridization of localized 4f levels with orbitals centered on surrounding atoms.
In this work, we will focus on the hybridization function, since as we shall see, it naturally gives information about the degree of localization.
In addition, the hybridization function
enables estimates of the Kondo temperature, via the effective coupling  $J_{\rm K}$ between f- and valence- states.\cite{fujita,wills}
Properties normally observed in Ce compounds,
such as the RKKY exchange interaction, Kondo singlet formation, and valence fluctuations,
are normally associated with $J_{\rm K}$.
For small values, the 4f electrons are essentially localized and can develop interatomic exchange via the RKKY coupling.
For increasing values of $J_{\rm K}$ the Kondo effect becomes dominant, so that a singlet many-body state develops.
Even larger values of the hybridization function push Ce systems into an itinerant regime, with fully delocalized bands.\cite{borje2,anna2}

The decisive parameters for $J_{\rm K}$ are the position of the 4f level with respect to the Fermi level
and the strength of the hybridization between 4f states and the remaining valence electrons.
Both these properties are available from {\it ab-initio} electronic structure theory, the first from calculations of the valence stability, using the Born-Haber cycle, see Ref.~\onlinecite{borje} and \onlinecite{Delin:97},
and the second from the hybridization function.
As a result of the different values of $J_{\rm K}$,
a plethora of characteristic behaviors have been identified for Ce compounds
where the starting point of the electronic structure is that of a localized 4f shell
that interacts more or less strongly with surrounding electron states.
For example, materials with dominant RKKY interaction are generally anti-ferromagnets,
although exceptions to this exist,
\eg CePdSb\cite{cepdsb} and CeRu$_2$Ge$_2$\cite{ceru2ge2} that both show a ferromagnetic ordering.
CePtIn and CeNiIn seem to have stronger
hybridization, and have been characterized as so called {\em dense} Kondo system,
without any observed magnetic ordering down to low temperatures~\cite{ceptin}.
Furthermore, Ce$_3$Bi$_4$Pt$_3$ is also a Kondo system,
although for this material a hybridization gap has been observed.\cite{ce3bi4pt3}
In addition to these effects, the interesting
phenomena of heavy fermion superconductivity has been observed, \eg in CeCu$_2$Si$_2$, CeCoIn$_5$ and CeRu$_2$Si$_2$.\cite{steglich}
Among the most strongly hybridized itinerant 4f materials, one finds, \eg $\alpha$-Ce, CeRh$_3$, CeN and CeFe$_2$.\cite{borje2,Weschke:92,anna2,cefe2}

In the present work we attempt to correlate known experimental characteristics of the electronic structure of a large set of Ce compounds (366 to be exact)
with information obtained from electronic structure theory, \ie the hybridization function, in order to distinguish between localized and delocalized 4f electron behavior. For a few of the investigated materials, where we identify the 4f shell to be only weakly hybridized, we have also calculated $J_{\rm K}$ and compared the calculated values to experimental data of the Kondo temperature $T_{\rm K}$.
One of the current trends in electronic structure theory is the efficiency and reliability of theory~\cite{Lejaeghere:16}
combined with the ability to generate large databases of electronic structure and related information
as demonstrated, for example, in Ref.~\onlinecite{carlos}.
The present work shows that this 
informatics approach is useful when combined with concepts from many-body model Hamiltonians,
and that predictions can be made 
for complex phenomena of correlated electron physics without actually solving the many-body problem itself.

\section{Methods}\label{sec:methods}
The electronic structure of 366 binary cubic Ce compounds has been investigated
within density functional theory (DFT) calculations using a full potential linear muffin-tin
orbital (FPLMTO) approach as implemented in the RSPt code.\cite{RSPt}
The data generated for this study will be incorporated in the IMS database for 4f and 5f systems.\cite{IMS}
If not stated otherwise, the structural input data
has been extracted from the Inorganic Crystal Structure Database (ICSD)~\cite{icsd} using cif2cell\cite{cif2cell}.
Alloys and systems containing deuterium or tritium, which 
are also present in the ICSD,
as well as systems with  large unit cells ($>$ 50 atoms), have not been included in our investigations.
All calculations of the basic electronic structure have been performed within the generalized gradient
approximation (GGA) in the formulation of Armiento and Mattsson (AM05)~\cite{AM05,Mattsson:08}.
We have neglected spin-polarization so that all calculations are directly comparable.
For Ce mono-pnictides, the maximum change in hybridization function strength is less than 10\% between the  nonmagnetic and the ferromagnetic calculation
(for details see the supplemental material, Fig. S1).
Spin-orbit coupling was neglected in this investigation for the sake of simplicity;
this approximation is not expected to have significant impact on the
conclusions of this investigation, and 
the techniques proposed here can be readily generalized to include spin-orbit coupling when necessary (\eg U and Pu systems).
We discussion the uncertainties arising from these assumptions in the supplemental material.
All calculations have been performed on the same level of accuracy (GGA DFT), important to identifying 
trends over large sets of data. 

We used the length of a reciprocal lattice vector divided by 0.15 to define the dimension of the Brillouin zone mesh 
in all calculations.
For comparison, a number of binary and ternary reference systems with different geometries has been taken into account,
using the same convergence parameters and $k$-point densities.
The hybridization function $\Delta (E)$, used to classify the compounds,
is obtained from an additional
iteration of the converged RSPt calculation, 
using an energy point mesh of 1501 points and a Fermi smearing of 1\,mRy.
Data from the f-electron database was used as a guide in assessing the results of our calculations.

\section{Results}\label{sec:results}
\subsection{The hybridization function}\label{sec:hybfunc}
\begin{figure}[tbh]
\centering
\includegraphics[width=.75\textwidth]{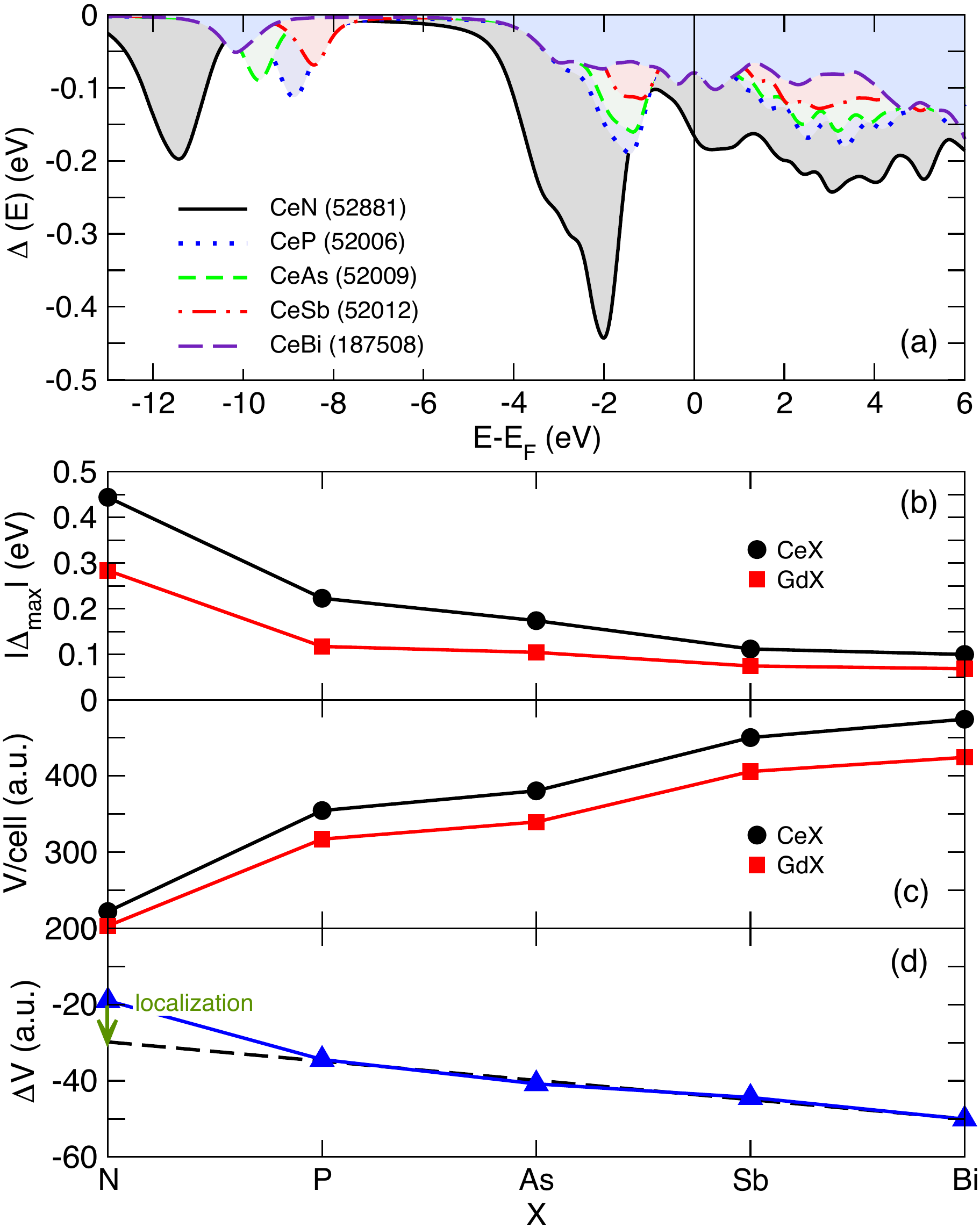}
\caption{(a) The calculated hybridization function $\Delta (E)$
as a function of the energy relative to the Fermi energy $E_{\rm F}$ for a series of Ce mono-pnictides. The values in the brackets denote the ICSD identification numbers~\cite{icsd}. 
(b) the maxima of $\Delta(E)$ for the systems shown in (a) (circles), compared to the values obtained for Gd pnictides (squares). 
(c) The experimental volume per unit cell of the same set of systems taken from the ICS database~\cite{icsd}.  
(d) The difference $\Delta V=V(CeX)-V(GdX)$ between the experimental volumes of CeX and GdX (X = N, P, As, Sb, Bi). 
The dashed black line denotes a linear fit of $\Delta V$(CeX) excluding N systems. 
The difference between the actual volume V(CeN) and the fit 
reveals the itinerant character of CeN. \label{fig:hyb-pnic}}
\end{figure}
In a quantum impurity model 
(Anderson) the energy-dependent hybridization function $\Delta(E)$
defines the properties of the bath surrounding the impurity cluster.  
The hybridization function describes the interaction of an impurity electron 
-- in our case the 4f electron of Ce -- 
with the bath consisting of all other electrons. 
With $G_0$ being the site projected Green's function calculated from density functional theory, 
and $H$ being the hybridization-free impurity Hamiltonian, with corresponding energy $E^{QI}$, 
we obtain, via the Dyson equation, an implicit expression for the hybridization function, $\Delta (E)$, as
\begin{equation}
{\rm G}^{-1}_{0} \,=\,(E - E^{\rm QI} )\,-\, \Delta(E) \,=\, (E-H)\,-\, \Delta(E).
\end{equation}
In the quantum impurity Anderson model, the hybridization function can be viewed as a measure of the tendency for Ce 4f band formation.
The larger the hybridization function, the bigger the overlap of the 4f orbital with all other orbitals, 
which is the principal indicator of itineracy.

The calculated hybridization function of a large set of Ce compounds is presented below, in a way that groups compounds naturally. 
In Fig.\,\ref{fig:hyb-pnic}(a) we show data for the well-known Ce monopnictides, CeX (X = N, P, As, Sb, Bi), that form in the NaCl structure. 
It may be observed that CeN shows a distinct peak about 2\,eV below the Fermi level 
and the area under the curve (shaded in gray) up to $E_{\rm F}$ is four times larger for CeN than for CeBi or CeSb, 
both of which exhibit a broad but flat $\Delta(E)$ curve.
The $\Delta(E)$ of CeP and CeAs lie in-between these extremes, reflecting the fact that the 4f states are neither as localized as in CeBi nor as delocalized as in CeN. 
Instead, the 4f electron orbitals overlap partially with the valence electrons of the ligand atoms. 
This trend is in perfect accord with established information about these compounds, discussed \eg in Ref.\,\onlinecite{Keis:63}, 
We conclude that for the Ce monopnictides, the hybridization function is a reliable measure of localization. 
Similar trends are observed when Ce is replaced by Gd (see Fig.\,\ref{fig:hyb-pnic}(b)), 
although the absolute value of the hybridization is strongly reduced for Gd compounds, 
with the maximum value ($E<E_{\rm F}$) being only 65\% of the hybridization of the corresponding Ce monopnictides. 
This is consistent with localized electron behavior in the Gd compounds (see Fig.\,S2 in the supplemental material).

Furthermore, for Ce monopnictides it has been found that the increasing localization of the 4f electron 
is correlated with the decreasing lattice constant,\cite{Litsarev:12} 
which is natural since a larger distance between Ce and neighbouring elements will lead to a reduced interaction.
In Fig.\ref{fig:hyb-pnic}(c) this trend is obvious for the Ce pnictides. 
As one goes down the group, the volume 
increases, whereas the hybridization function drops
by a factor of 4-5 (see Fig.~\ref{fig:hyb-pnic}(b)). 
To understand this trend in detail, it is relevant to make a comparison with compounds that are known to have completely localized 4f states,
such as the Gd pnictides. 

A comparison of the experimental volumes of Ce mono-pnictides CeX (X = N, P, As, Sb, Bi) 
with those of the corresponding Gd systems confirms the itinerant character of CeN.
For both Gd- and Ce-based pnictides, the volume shows some irregularities when different ligand atoms from group 15 are considered, 
with an overall trend of increasing volume for the heavier ligands (Fig.\,\ref{fig:hyb-pnic}(c)).
This trend reflects to some degree the change in the nature of the 4f states,
but also the difference in atomic size of the group 15 elements. 
In order to isolate the effect coming from the 4f states, 
we compare the volume of the Ce-pnictides with the volume of the corresponding Gd-pnictides. 

It is well known that the latter rare-earth element is trivalent with a fully localized, non-hybridizing 4f shell. 
The difference in volume between these sets of compounds is shown in Fig.\ref{fig:hyb-pnic}(d). 
Since the lanthanide contraction is known to result in smaller volumes of heavier rare-earth elements, 
the difference in volume between CeBi and GdBi may be seen as a refection of this fact. 
If the 4f electrons in the remaining Ce-pnictides were completely localized, giving rise to trivalent and chemically inert 4f states, 
the difference in volume of CeBi and GdBi would be exactly the same for all compounds plotted in  Fig.\ref{fig:hyb-pnic}(d). 
The deviation from this behavior signals the increased interaction between the 4f states and the ligand orbitals as one moves from heavier to lighter elements of group 15, 
and it is seen that the deviation behaves almost linearly.
The deviation of the $\Delta V$ curve from the linear behavior is illustrated by the difference of the true values and the extrapolated values,
 marked by a dashed line in Fig.\ref{fig:hyb-pnic}(c). 
This deviation is due to a drastic change in the electronic structure of CeN, compared to the other compounds. 
In short, this marked deviation is caused by the itinerant character of the 4f electron in CeN, 
a fact that is also clear from the hybridization function that clearly is largest for this compound.

The data in Fig.\ref{fig:hyb-pnic} suggest that the hybridization function is a good measure when trying to identify general trends of the electronic structure of Ce-pnictides. To
investigate whether this observation holds also for other Ce compounds, the hybridization function has been calculated for a large set of known Ce compounds, with different crystal symmetry and composition.
The hybridization functions of CeRh$_3$ and CePt$_3$ are in these calculations found to
possess distinct peaks at 1.21\,eV and 2.53\,eV below the Fermi level (Fig.\,S3).  The largest values $|\Delta(E)_{\rm max}|$ lie in the range of $\sim$ 0.5-0.7 eV, while the majority of the compounds have smaller
hybridization function magnitudes (cf. Table\,\ref{tab:delta}, where we list peak values of several considered compounds).  CeN, CeRh$_3$ and CePt$_3$ are known to have essentially itinerant 4f electrons, which is consistent with the large values of $|\Delta(E)_{\rm max}|$.\cite{Weschke:92, Severin:94} In contrast, CeCoIn$_5$ is an example of a very weakly hybridized compound with a $\Delta(E)$ that
possesses weak, broad features without distinct peaks, (pink dash-dotted curve in Fig.\,S3). This implies
at a very pronounced localization of the 4f shell, as expected from the literature.\cite{Curro:01}  Table\,\ref{tab:delta} suggests that Ce$_3$Ge and CeCoIn$_5$ are the least hybridized compounds, with expected localized electron behavior of the 4f shell. The table also shows a large group with intermediate hybridization among these systems, as exhibited by \eg $\alpha$-Ce, CeRu$_2$Si$_2$, CePtIn, Ce$_3$Bi$_4$Pt$_3$, CeNi$_2$, CePt$_2$ and CePt$_5$. Experimental data for CeRu$_2$Si$_2$, a well known heavy fermion material, is consistent with its position as an intermediately hybridized material.

The discussion above suggests that $\Delta(E)$ can be viewed as a measure
of the degree of localization of the Ce 4f electron and can therefore be used to classify large groups of Ce compounds in terms of the degree of localization. However, since
our goal is to compare $\Delta(E)$ for many systems, the full energy dependent hybridization function may not be a practical gauge. As an alternative we explore the maximum value of $\Delta(E)$ for $E\leq E_{\rm F}$, (see Tab.\,\ref{tab:delta} and  Fig.\ref{fig:hyb-pnic}(a) for Ce and Gd mono-pnictides).
An alternative choice is to integrate over $\Delta(E)$ using
 $|\Delta_{\rm int}|(E_{\rm F})=\int_{-\infty}^{E_{\rm F}}\Delta (E)dE$
as a measure, cf. Tab.\,\ref{tab:delta}. We will see later that 
how we choose to compare the hybridization function has some influence on the grouping of the systems, depending on the degree of localization, but the overall trends are picked up by any of the
choices discussed here.
From Tab.\,\ref{tab:delta} we note that $|\Delta_{\rm int}| (E_{\rm F})$ seems to give a better representation of the level of itineracy when compared to experimental findings. For instance, this measure puts CeNi$_2$ on
the more itinerant side, which is consistent with the analysis of \eg Ref.\,\onlinecite{ceru2ge2}. However, in order to use data-mining algorithms on a large body of compounds
to identify trends in the degree of localization-delocalization of the 4f shell, one can use any of the forms of the hybridization interaction discussed in this paper.

\begin{table}[bhtp]
\caption{Extrema of the 4f hybridization function (in absolute values) $|\Delta_{\rm max}|$  and  integral over the hybridization function up to the Fermi level  $|\Delta_{\rm int}|(E_{\rm F})$ for selected reference systems. Note that the units of $|\Delta_{\rm max}|$  and  $|\Delta_{\rm int}|(E_{\rm F})$ are different,  $|\Delta_{\rm max}|$ is given in eV and  $|\Delta_{\rm int}|(E_{\rm F})$ in eV$^2$.}
\begin{center}
\begin{tabular}{L{1.9cm}L{1.9cm}L{1.8cm}C{1.8cm}C{1.8cm}}\hline\hline
System &Crystal &Space &$|\Delta_{\rm max}|$&$|\Delta_{\rm int}|(E_{\rm F})$\\
 &structure  &group &$ (E\leq E_{\rm F})$&  \\\hline
Ce ($\alpha$) &cubic &Fm-3m&0.278& 0.518\\
Ce$_3$Ga&cubic &Pm-3m&0.073& 0.090\\
Ce$_3$Ga&cubic &Pm-3m&0.073& 0.090\\
CeCu$_2$Si$_2$&tetragonal &I4/mmm&0.111& 0.555\\
CeCoIn$_5$&tetragonal & P4/mmm &0.113& 0.338\\
CeRu$_2$Si$_2$&tetragonal &I4/mmm&0.135& 0.608\\
Ce$_3$Bi$_4$Pt$_3$& cubic &I-43d &0.162& 0.543\\
CePt$_5$& hexagonal &P6/mmm&0.172&0.696\\
CePt$_2$& cubic &Fd-3mS &0.186&0.771\\
CePtIn& hexagonal &P-62m&0.235&0.559\\
CeNi$_2$& cubic &Fd 3m S	&0.283&0.903\\
Ce$_2$O$_3$&trigonal&P-3m1&0.629 &1.428\\
CePt$_3$&cubic &Pm-3m&0.569&1.554\\
CeRh$_3$&cubic &Pm-3m&0.674&1.799\\\hline\hline
\end{tabular}
\end{center}
\label{tab:delta}
\end{table}%

\subsection{Trends of the hybridization function in cubic binary Ce compounds}\label{sec:cubbinsys}
\begin{figure}[tbh]
\centering
\includegraphics[width=.85\textwidth]{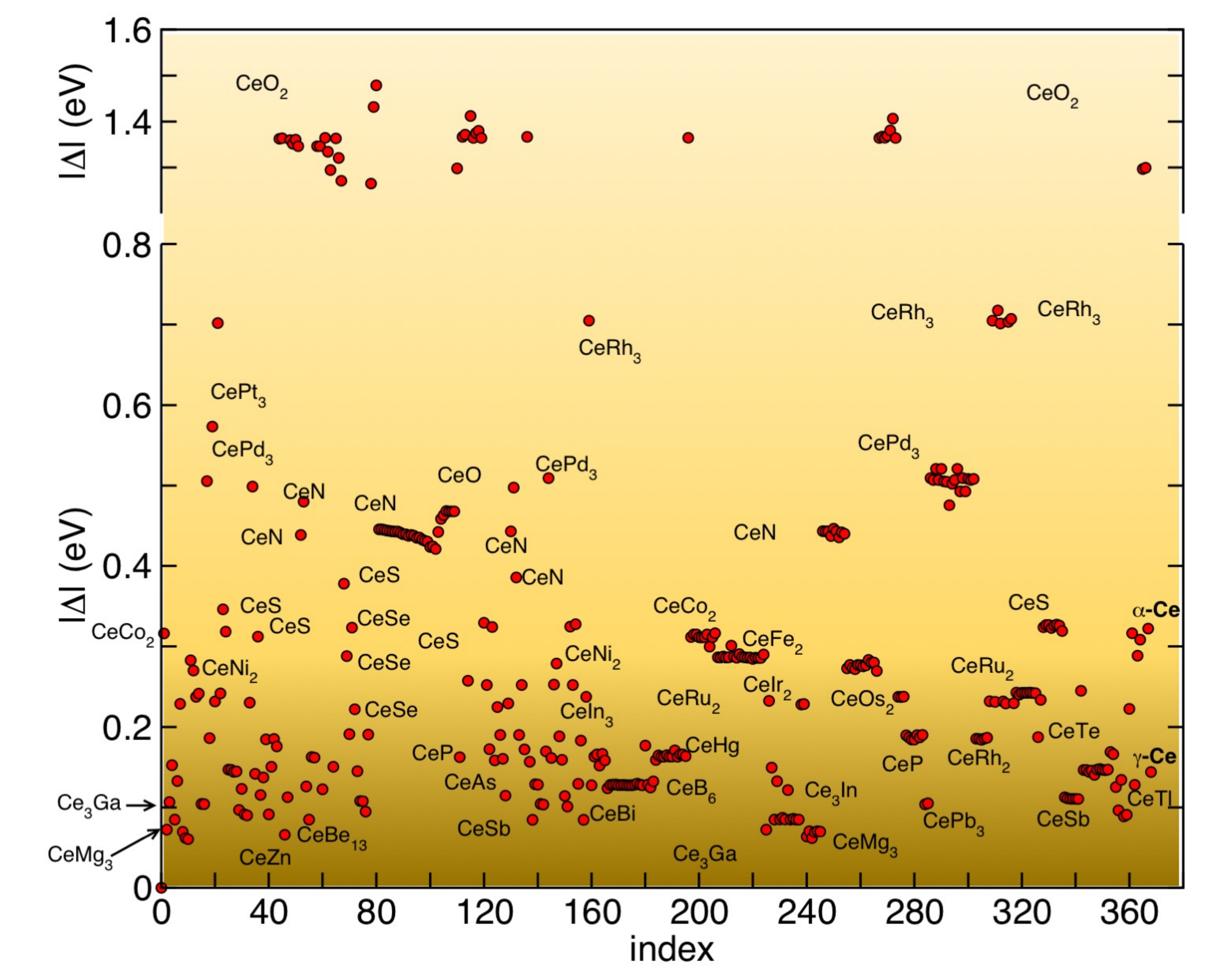}
\caption{Extrema of the hybridization function (in absolute values)  $|\Delta (E)|$ for 366 binary cubic Ce compounds. The color scale indicates the degree of localization versus itineracy, where darker yellow color represents increased level of localization and lighter yellow color represents increased itineracy. The index on the x-axis just numbers the systems and has no physical meaning. \label{fig:hyb-all}}
\end{figure}

We have shown, for a representative subset of systems, that the hybridzidation function is a reliable criterion
to distinguish between
localized and de-localized 4f systems, even at a rather simple level of approximation.
$\Delta(E)$ has been calculated using plain DFT (GGA) for the known cubic binary compounds (regarding the limits mentioned in Sec.\ref{sec:methods}, 366, according to ICSD\cite{icsd}). Instead of plotting the hybridization function over the whole energy range, the extrema of it are given in Fig.\,\ref{fig:hyb-all}, evaluated for occupied states. Even though there is a certain spread in the data points due to differences in lattice parameters, caused by differences in the experimental techniques and conditions extracting them (such as temperature and external pressure), the trend observed for the systems discussed in the previous sections seems to carry over to all investigated cubic binary Ce compounds. Fig.~\ref{fig:hyb-all} shows that the hybridization function
is spread over a large interval, and that compounds that are well established as either localized or itinerant, naturally find their place
in regions with low and high hybridization, respectively.
The transition region is drawn in the figure
so as to not be sharp and covers elements that normally are associated with pronounced Kondo or mixed valence behaviour.

We now make a connection to traditional measures of the level of itineracy; the 4f bandwidth and its relation to the Coulomb U.
Cubic CeO$_2$ has, in LDA theory, a very broad 4f band which strongly hybridizes with the O p orbitals (cf. Fig.~S3(a)). For CeN, the pnictide with the most itinerant character, the picture is very similar, except that there is no band gap (see Fig.~S3(b)). However, even though the width of the 4f band is very similar to that of CeO$_2$, the overlap between Ce f and N p is
small compared to the oxide. Going down the N group (no. 15) the width of the 4f band, as obtained from LDA theory, decreases (as can be seen from Figs. S3(c) and (d)) and the tendency for electron localization becomes stronger. In fact,
only  the 4f DOS of CeN, CeRh$_3$ and CeO$_2$ have widths that are comparable or larger than the Hubbard U (which is in the order of  2-4 eV for itinerant electron systems, and 4-7 eV for localized electron systems, with precise values depending on the method of calculation and the compound, see Refs.\,\onlinecite{Topsakal:14, Banik:14,Jiang:05,Boring:92}), which according to conventional criteria would render only these three compounds as being itinerant.
The values of $|\Delta(E)|$ give entirely consistent conclusions regarding
distinguishing itinerant 4f compounds from
those with 4f localization. We conclude that the hybridization function is a  simple measure for
quantifying the degree of localization of the 4f electrons in Ce compounds, and that it is
 reliable 
for high throughput screening investigations. It also provides important parameters for many-electron physics, and establishes a coupling between {\it ab initio} electronic structure theory and many-body model Hamiltonians, in this case the single impurity Anderson Hamiltonian.

\subsection{Correlation between hybridization and lattice parameter}\label{sec:correlation}
The goal of the present paper is not only to classify Ce compounds regarding their itinerant character but to link this information to other tuneable quantities.
If this link can be established with a sufficient accuracy it will
provide a way to design systems with desired electronic properties
and serve as a way to control the degree of correlation.

\begin{figure}[tbh]
\centering
\includegraphics[width=.9\textwidth]{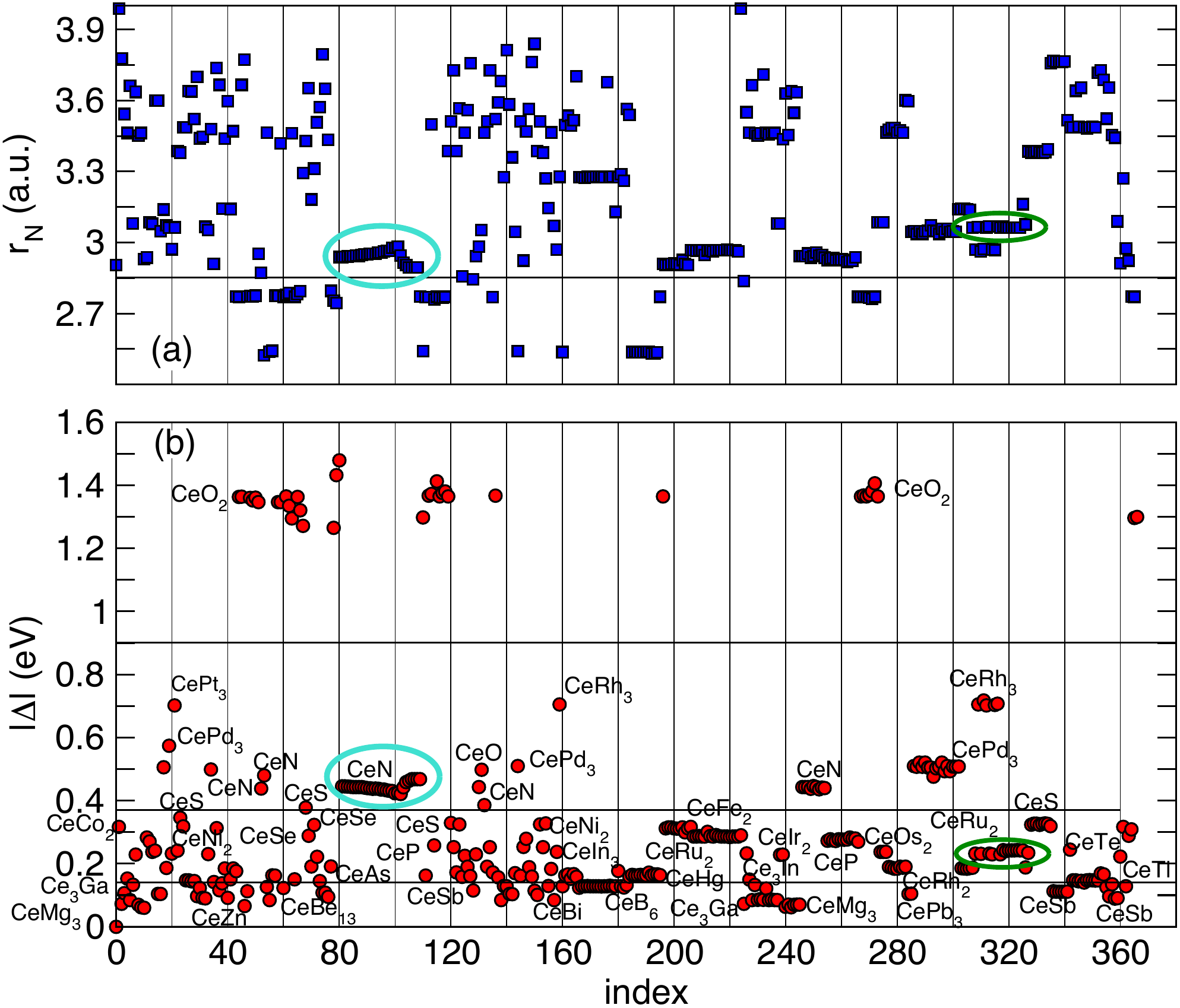}
\caption{The Wigner-Seitz radius ($r_N$) of the Ce atom
at the experimental volume for cubic binary Ce compounds (a) and the extrema of the hybridization function (absolute value)
$|\Delta(E)| (E\leq E_{\rm F})$ for the same set of compounds (b). The circles areas mark sets of CeN and CeRu$_2$ systems indicating an inverse relation between the volume and the size of the hybridization. For more details see text.\label{fig:vn-hyb}}
\end{figure}

The discussion of the mono-pnictides in Sec.\,\ref{sec:hybfunc} has shown that the lattice parameter increases in the same steplike manner as the localization decreases
(see Figs.~1(a) and (b)). Therefore, the lattice parameter seems to be a suitable quantity to link
to the hybridization function. This is supported by the fact
that $|\Delta(E)|$ is correlated with the bond distance,
as is clear from systems where the lattice constant of a certain compound
in different experiments is found to be slightly different
(presumably these differences may be due to differences in sample purity or to differences in temperature or pressure for the experimental investigation).
Examples where this is seen are found for CeN or CeRu$_2$, as is clear from the circled areas of Fig.~\ref{fig:vn-hyb}(a) and (b).

The Wigner-Seitz radius for all
compounds investigated is given in Fig.~\ref{fig:vn-hyb} (a). For comparison the
corresponding hybridization extrema are also plotted (Fig.~\ref{fig:vn-hyb}(b)). Since the composition of the systems and the alloy partners are very diverse, the radii
are spread over a broad region from 2.5 to 4.0\,a.u.
However, one can distinguish
trends such as the very small volumes of the CeO$_2$ systems ($r_N \approx 2.75$~a.u.) where the hybridization function is large, and the large $r_N$ of CeMg$_3$
which has one of the smallest $|\Delta(E)_{\rm max}|$ values, (cf. Fig~\ref{fig:vn-hyb}).

\begin{figure}[tbh]
\centering
\includegraphics[width=.9\textwidth]{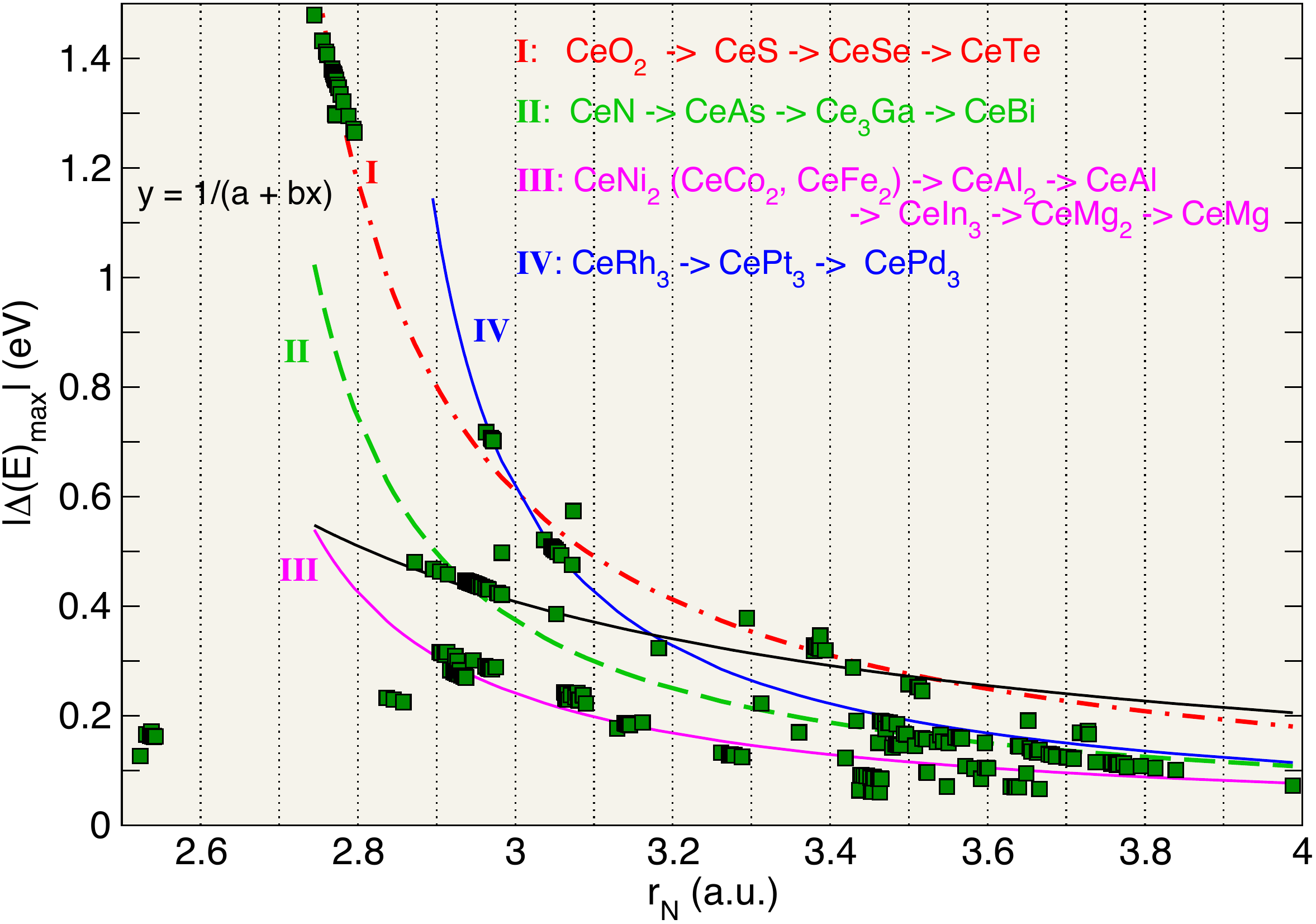}
\caption{Scatter plot of the absolute values of the hybridization function $|\Delta{E}_{max}|$ versus the radius of the volume per atom $r_N$. The lines are
inverse linear fits. The solid black line represents the fit over all data sets except the ones for CeB$_6$ and CeBe$_{13}$. The two systems have been excluded since Ce is basically an impurity in a boron or berylium matrix and   $r_N$ is mostly determined by the light elements instead by Ce.\label{fig:crosscorr1}}
\end{figure}

Visual inspection of Fig.~\ref{fig:crosscorr1}, where $|\Delta(E)_{max}|$ is directly plotted versus  $r_N$, reveals, at least qualitatively,
a correlation between the two quantities. The hybridization function decays basically
as $1/(a + b\cdot r_N)$, where $a$ and $b$ are fitting parameters.
The solid black line in Fig.~\ref{fig:crosscorr1} represents the correlation between the hybridization function and
the Wigner-Seitz radius.
A further inspection of Fig.~\ref{fig:crosscorr1} reveals that the overall behavior for all systems is the same but the slope can be different for different classes of systems.
CeO$_2$ seems to form one group including also some CeZ (Z = S, Se, Te) (cf. red dashed-dotted line (I)  in Fig.~\ref{fig:crosscorr1}). Another group (II) is formed by the mono-pnictides (green dashed line in Fig.~\ref{fig:crosscorr1}). Systems of the type CeM$_2$ (M = 3d transition metal) decay with  a different slope (III) and can be viewed as a third group together with Ce compounds with light metals such as Al.  The CeT$_3$ (T = Pt, Pd, Rh) systems do not fit in any of these groups indicated by the blue line in  Fig.~\ref{fig:crosscorr1}.  This suggests that besides the obvious decay of
$|\Delta(E)_{\rm max}|$ with $r_N$ there might be a more subtle dependence which determines the slope.
This brings up the question of the cause of
these different slopes, i.e. the fine-structure in the $1/r_N$ decay.  The
groups defined above differ in
electronic structure, \ie the type of valence electrons of the ligand atom, their parity, and the width of these ligand valence bands,
which are in general known to be  important factors for the hybridization. CeO$_2$ as well as CeZ (Z = S, Se, Te)  have nominally p$^4$ valence electrons from the non-Ce element and they have the same parity as the Ce f electron. Therefore, these electrons are expected to hybridize which is reflected in the large values of the hybridization function. Even though the $r_N$ of the CeZ is already quite large the corresponding $|\Delta(E)_{\rm max}|$ lies clearly above all other systems with similar volume, see Fig.\,\ref{fig:crosscorr1}.  With decreasing number of p electrons the slope increases and the overall size of the hybridization becomes  smaller as can be seen for the p$^3$ systems which contain the mono-pnictides (green dashed line). Systems without a significant p contribution to the valence electrons form the lowest group (pink curve in Fig.\,\ref{fig:crosscorr1}). This class has basically s or 3d  valence electrons which have opposite parity than the f electrons and consequently the tendency to hybridize with the f type electrons is much lower. The fourth group is special since the T elements (T = Pt, Pd, Rh) also possess mainly d type valence electrons but at the same time they have comparable large hybridization energies. One difference between the CeT$_3$ and the CeM$_2$ compounds is that due to 3(4)d core electrons the outer d electrons form broader bands~\cite{Sigalas:92} which then leads to a larger hybridization with the f electron. Besides the different core structure these compounds have three 4d or 5d atoms per formula which multiplies the effect.  CeT$_2$ (T = Pt, Pd, Rh) do not show the same behavior. Their $|\Delta(E)_{\rm max}|$ values are much smaller and they fit in the 3d group. Now one might think that CeT$_x$ compounds with larger $x$ values would lead to even higher hybridization energies, but, for example, the  $r_N$  of CePt$_5$ amounts to 3.08\,\AA~ but its hybridization is
0.172\,eV, which is smaller
than the one obtained for CePt$_3$ (see Tab.\,\ref{tab:delta}). However, CePt$_5$ crystallizes in a hexagonal structure (space group no. 191) and the different geometrical arrangement has certainly an influence on the hybridization of the orbitals.

Summarizing our results so far, we observe a clear correlation between the size of the hybridization function and the Wigner-Seitz radius. In addition, we discovered a fine structure within the Ce compounds, \ie the slope of the decay is also determined by the electronic configuration of the non-Ce element.  Together these two findings might be used for the search of  Ce systems with tailored 4f character and functionality, or as a quantitative characterization tool as regards the degree of itineracy versus localization of known compounds. Overall the dependence of the hybridization function on the Wigner-Seitz radius (or volume) gives a first indication how to tune the volume to obtain a more itinerant or localized electron system. The importance of the 4f hybridization for materials properties  of correlated materials has already been pointed out in an earlier work by Koelling et al.\cite{Koelling:85} This is relevant \eg in high-pressure science since it gives information on which compressions are needed in order to increase the itinerant character with a certain amount. For additional fine tuning of the degree of localization/itinerancy, the type of valence electrons, and crystal geometry, are also relevant parameters. Since computations of the type reported here are expedient, this makes theoretical investigations of the hybridization function extremely useful as a precursor to many experimental investigations of correlated electronic structures of Ce compounds (and possibly other f-electron based materials).


To confirm
this correlation quantitatively, a statistical analysis of the
$r_N(i)$ and $\Delta(E)_{\rm max}(i)$
data sets has been performed, where $i$ is the sample index (for simplicity we skip the argument $E$ in the following).
We use Pearson's correlation coefficient $\rho(r_N,\Delta)$\,\cite{Pearson:1895,Stigler:89} and mutual information theory $I(r_N,\Delta)$ in order to quantitatively estimate the correlated or anti-correlated nature of the two data sets.
In Fig.~\ref{fig:crosscorr1}(a) and (b), we discussed the distribution of 366 Ce based compounds in the radius and corresponding hybridization energy range.
From this information a  histogram distribution was created using 25 bins for both
ranges of volume/atom (2.5\,\AA$^3$ to 4\,\AA$^3$) and hybridization energy (0.0 to -1.5\,eV). The probability of the $i$-th bin for a $r_N$ data set is obtained using
\begin{equation}
P_{r_N}(i)=\frac{n_{r_N}(i)}{L},
\end{equation}
where $n_{r_N}(i)$ is the number of compounds in the $i$-th bin, and $L$ is the total number of compounds in the volume data set.
Similarly, the probability of the $j$-th bin in the hybridization data set is given by
\begin{equation}
P_{\Delta}(j)=\frac{n_{\Delta}(j)}{L}.
\end{equation}

\begin{figure}[tbh]
\centering
\includegraphics[width=.99\textwidth]{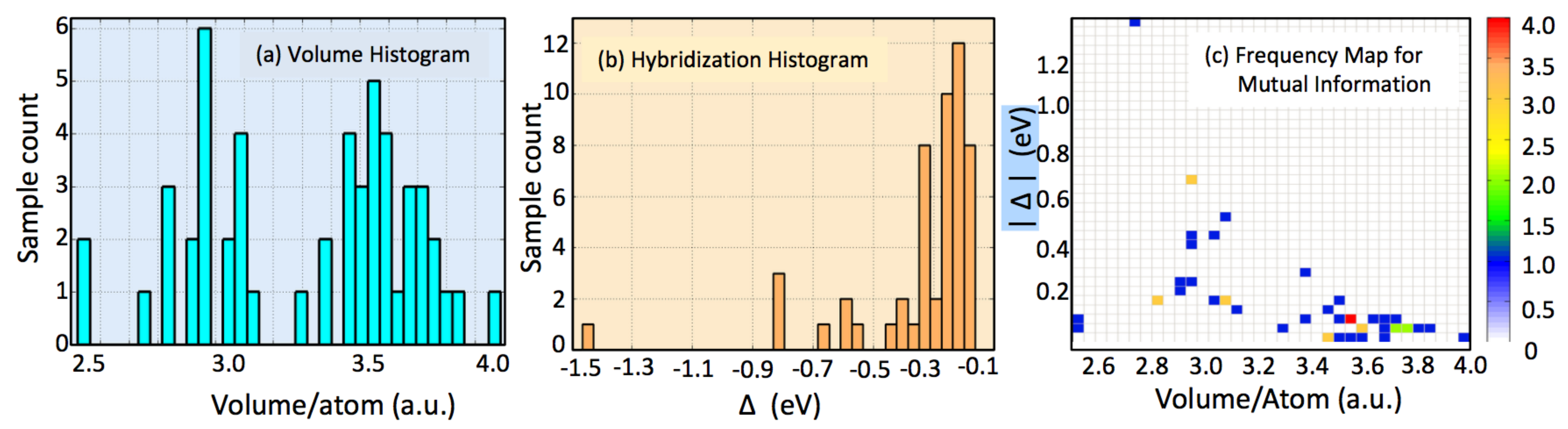}
\caption{Histograms in (a) and (b) are the distribution of 52 Ce compound data points on volume/atom ($r_N$) and hybridization energy ($\Delta_{\rm max}$)
correspondingly. (c) Two-dimensional frequency map obtained from {Eqn.~\ref{eq:mutualinfo}} where colors represent the number-count of Ce compounds in each cell.  Data are shown for the $L = 52$ data set which contains only unique systems. The color bar quantifies the color schemes
used in this plot. The frequency map provides joint probability distribution and measures mutual information $I(r_N,\Delta)$ (see text for details). \label{fig:crosscorr2}}
\end{figure}

The correlation coefficient which corresponds to the scatter plot of $r_n(i)$ versus $\Delta(i)$ shown in Fig.\,\ref{fig:crosscorr2} can be determined by using
\begin{equation}
\rho(r_N,\Delta)=\frac{\sum_{i=1}^{L} (r_N(i)-\bar{r}_N)(\Delta(i)-\bar{\Delta})}{\sqrt{(\sum_{i=1}^{L} (r_N(i)-\bar{r}_N)^2)
(\sum_{i=1}^{L} (\Delta(i)-\bar{\Delta})^2)}}.
\end{equation}

Here $\bar{r}_N$ and $\bar{\Delta}$ are the average radius/atom and average hybridization for the $L$ Ce-compound data set. Though we have performed calculations for 366 data sets the actual number of different compounds is 52 because many systems having  the same chemical formula been investigated under different conditions, see Fig.\,\ref{fig:crosscorr2} (a) and (b).
In order to predict materials properties, the smaller data set is more suitable since it contains only composition dependent information. We have also performed the analysis for the larger set which includes information on the volume dependence of single compounds. The latter value will be given in parentheses.
We find a correlation coefficient of $\rho(r_N,\Delta) = -0.54 (-0.56)$. Since $\rho$ varies, by definition, between 1 (complete correlation), 0 (no correlation), and -1 (maximal anti-correlation), this clearly indicates that the correlation between our two properties,  $r_N$ and $\Delta$, is at the border between moderate and strong anti-correlation.

In Fig.~\ref{fig:crosscorr2}c, we plot the two-dimensional frequency-map of $r_N(i)$ and $\Delta(i)$ data sets. The color scheme indicates the population of Ce compounds in each cell.
The heat-map provides information on the joint probability distribution $P(i,j) = n(i,j)/L^2$, where $n(i,j)$ is the number count in the $i,j$-th cell, and $L$=52 (366)  is the total size of the data set. This quantity is essential to determine the {\em mutual information}.
Unlike the correlation coefficient, the mutual information is a positive definite quantity which provides a measure of the {\em information entropy}\cite{Shannon:49} , and therefore, quantifies the amount of information that can be achieved for one random variable through another set of random variables. Formally, the mutual information is defined as
\begin{equation}\label{eq:mutualinfo}
I(r_N,\Delta)=\sum_{i\in r_N,j\in \Delta} P(i,j) \log\!\left(\frac{P(i,j)}{P(i)P(j)}\right),
\end{equation}
where $i$ and $j$ indices stand for the $i$-th row $j$-th column ($i,j$-th cell) in the frequency map.
For our given data set of $r_N$ and $\Delta$,
the mutual information is $I(r_N,\Delta)=$ 0.42 (0.73). This value of $I(r_N,\Delta)$ clearly demonstrates a quite high predictive
value for
electron-electron correlation from only volumetric data and vice versa, and thus opens up pathways for machine aided new materials design principles. The value for the large data set points to an even higher predictive power which basically reflects the fact that the localization within a material is strongly related to the volume.
\subsection{Kondo couplings and Kondo temperature}\label{sec:kondo}
\begin{figure}[tbh]
\centering
\includegraphics[width=.85\textwidth]{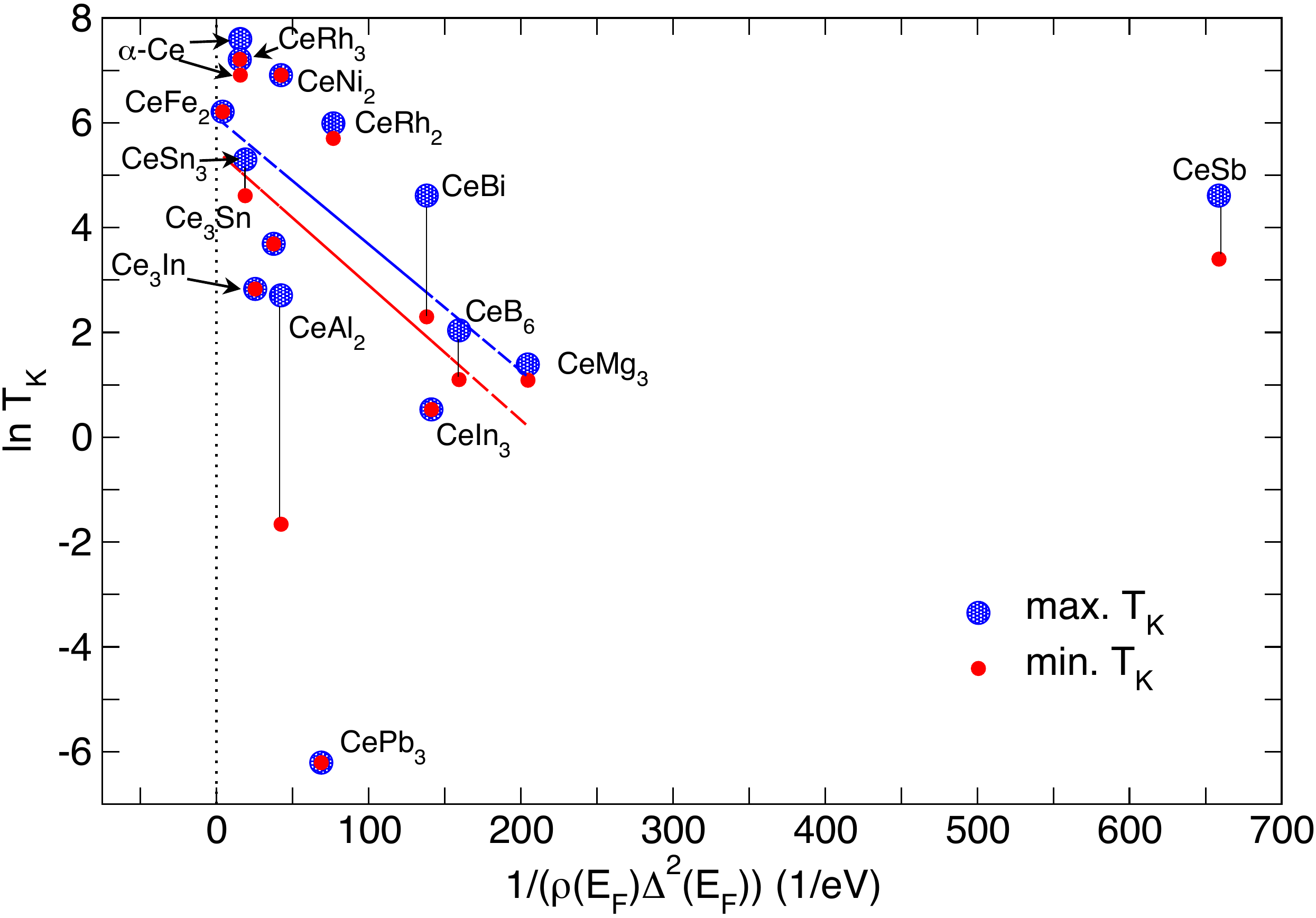}
\caption{Experimental Kondo temperatures $T_{\rm K}$, obtained from literature for known cubic binary Ce compounds, as function of the calculated hybridization function $\Delta(E_{\rm F})$ and DOS(E$_F$).  Both calculated values are taken at the Fermi level $E_{\rm F}$. The blue (red) line is a linear fit through the largest (smallest) $T_{\rm K}$ value found for a given compound. The vertical black lines are guides to the eye connecting the $T_{\rm K}$ min and $T_{\rm K}$ max of a system.} \label{fig:TK}
\end{figure}

In the previous part of the paper we have argued that the hybridization function can be used as a measure for the degree of localization in a Ce-compound. However, weakly hybridized 4f electrons are known to give rise to the Kondo effect, where they form scattering centers for the other valence electrons. When varying temperature, this leads  to the experimentally observed minimum of the electrical conductivity. Determining the Kondo temperature is routinely done by addressing an impurity problem, either stand alone as in the Anderson impurity model\,\cite{Lacroix:77, Malterre:96} or within dynamical-mean-field theory\,\cite{Georges:96,Sakai:07}. However, the solution of the multi-orbital Anderson model is a demanding task and not suitable for big data analysis. Here our aim is to explore if it is possible to identify trends in the Kondo temperature of Ce-based compounds by looking only at quantities extracted from DFT investigations, \ie without explicitly solving the Kondo problem. 
Toward this goal we use the above discussed large data set generated for Ce compounds and search for systematic trends that reveal themselves in the aggregate. To assist in this analysis 
we make use of the formula derived by Lethuillier and Lacroix-Lyon-Caen for the single-orbital impurity model.~\cite{Lethuillier:78, Lacroix:77} In this approach the Kondo temperature of a Ce compound can be written as\cite{Lethuillier:78}
\begin{equation}\label{eq:TK}
T_{\rm K}\,=\, \alpha\, \exp\left[-\frac{2}{(2J+1)J_{\rm K}\,\rho(E_{\rm F})} \right]
\end{equation}
where $J=5/2$ for tripositive Ce ions and $\alpha$ is a proportionality factor. $\rho(E_{\rm F})$ is the DOS of the spd states at the Fermi level and is here obtained from a separate calculation with the Ce 4f states are treated as core electrons. In the flat band approximation\,\cite{Hewson:93},
the Kondo coupling parameter $J_{\rm K}$ is  given by
\begin{equation}\label{eq:JK}
J_{\rm K}\,=\, \frac{2\Delta^2(E_{\rm F})}{(E_{f}-E_{\rm F})}
\end{equation}
with $\Delta$ being the hybridization energy. The energy difference in the denominator is between  the 4f level and the Fermi niveau.
For simplicity the influence of crystal field effects is neglected  and we assume $E_{f}-E_{\rm F} \simeq -3$ eV which is a good approximation for Ce systems\,\cite{Delin:97}. $\Delta(E_{\rm F})$ is the hybridization function at the Fermi level, as discussed in the previous sections. Practically, the logarithm of Eq.\,\ref{eq:TK} is easier to analyze, and we proceed with the expression:
\begin{equation}\label{eq:lnTK}
\ln{T_{\rm K}}\,\sim\, \frac{1}{2\,\Delta^2(E_{\rm F})\,\rho(E_{\rm F})}.
\end{equation}
Plotting the experimental Kondo temperatures as a function of calculated data of the denominator of Eqn.\,\ref{eq:lnTK}, is expected to result in a  linear, decaying behavior. However, most experimental Kondo temperatures are given as a range, due to data provided from different experiments or different samples, \eg for CeBi 10-20\,K \cite{Sera:83}, and
according to Ref.\,\onlinecite{Kasuya:87} all Ce mono-pnictides should have Kondo temperatures between 50 and 100\,K.
Therefore, in Fig.\,\ref{fig:TK} we provide two data sets, one containing the smallest $T_{K}$  (small red circles in Fig.\,\ref{fig:TK}) found in literature and one for the largest (large blue circles in Fig.\,\ref{fig:TK}).
\begin{table}[htp]
\caption{Experimental Kondo temperatures $T_{\rm K}$, the calculated  hybridization energies $\Delta(E_{\rm F})$ (eV), the  corresponding calculated total DOS at the Fermi level  (with the 4f electron being treated as core electron) $\rho(E_{\rm F})$ (states/eV) are given for several cubic binary Ce compounds. The distance between the Fermi level $E_{\rm F}$ and the position of the 4f peak  $E_{f}$ (eV) is assumed to be 3 eV, according to Ref.\,\onlinecite{Delin:97}.}
\begin{center}
\begin{tabular}{L{1.3cm}L{1.3cm}C{1.3cm}C{1.3cm}R{1.5cm}R{1.5cm}R{1.5cm}}\hline\hline
System&ICSD& $|\Delta ({\rm E_F})|$ &$\rho(E_F)$ & $(\rho\Delta^2)^{-1}$ &$T_{\rm K}$ (max)& $T_{\rm K} $ (min)\\\hline
$\alpha$-Ce &41823 &  0.322 &0.617  & 15.64   & 2000\cite{Liu:92}&1000\cite{Zoefl:01}\\
CeAl$_2$ &606387     & 0.129    &  1.417  &     42.41& 15\cite{Patthey:90} &0.19\cite{Lacroix:77}\\
CeB$_6$ &612731&0.124&0.408&159.39&7.7\cite{Suga:14}&3\cite{Suga:14}\\
CeBi         &187508          & 0.079    &  1.160  &  138.13  & 100\cite{Kasuya:87} & 10\cite{Sera:83} \\
CeFe$_2$ &620998&0.226    & 4.963&3.94 &500\cite{Cho:03}&500\cite{Cho:03}\\
Ce$_3$In &621360&0.124     & 2.561&25.40&17\cite{Wang:10}&17\cite{Wang:10}\\
CeIn$_3$ &621361     & 0.086    &  0.957  &  141.24  & 1.7\cite{Lacroix:77} & 1.7\cite{Lacroix:77}\\
CeMg$_3$& 621498&0.062&1.271&204.68& 4\cite{Das:11}&3\cite{Das:11}\\
CeNi$_2$ &102229  & 0.174    &  0.779 &    42.40 &$>$1000 \cite{Weibel:95}&$>$1000\cite{Weibel:95} \\
CePb$_3$ &621777   & 0.104    &  1.343 &    68.84  &  0.002\cite{Lethuillier:78} & 0.002\cite{Lethuillier:78} \\
CePd$_3$ &107546& 0.054&0.262& 1308.91&350\cite{Weibel:95,Onuki:87}&240\cite{Weibel:95,Onuki:87}\\
CeRh$_2$ &621938  & 0.150   &   0.579 &   76.76& 400\cite{Sugawara:94}&$>$300\cite{Onuki:87}\\
CeRh$_3$ &604325& 0.174&2.152&15.35& 1350\cite{Weibel:95}&1350\cite{Weibel:95}\\
CeSb &52012             &  0.112   &  0.121& 658.84& 100\cite{Kasuya:87}&30\cite{Sera:83}\\
Ce$_3$Sn &622247&0.146      & 2.488&18.86&40\cite{Wang:10}&40\cite{Wang:10}\\
CeSn$_3$ &622224&  0.146&1.231&37.57&200\cite{Onuki:87}&100\cite{Onuki:87}\\\hline\hline
\end{tabular}
\end{center}
\label{tab:kondo}
\end{table}%
Fig.\,\ref{fig:TK} indeed reveals the expected linear behavior between the logarithm of the experimental Kondo temperatures and the calculated values of $(\rho(E_F)\Delta^2(E_{\rm F}))^{-1}$. The spread of the data in the figure is partially related to the different experimental Kondo temperatures, which are obtained from different measurement techniques, \eg from magnetic susceptibility measurements\,\cite{Das:11, Lethuillier:78}, photoelectron experiments\,\cite{Huefner:03} or trends that can be obtained from resistivity curves\,\cite{Hayashi:16}, in addition to differences in quality of the samples (single crystals\,\cite{Venturini:06}, polycrystalline samples\,\cite{Galera:85}, and nanocrystals\,\cite{Chen:07}). Other reasons for the spread are naturally connected to approximations made in the theoretical analysis. However, overall the expected trend is clearly reproduced. Heavy fermion systems such as CeB$_6$ and CeMg$_3$ provide large $(\rho(E_F)\Delta^2(E_F))^{-1}$ which are connected to small $T_K$.
With decreasing $(\rho(E_F)\Delta^2(E_F))^{-1}$, $\ln T_{\rm K}$ increases, 
\ie the heavy fermion character vanishes; and finally, the smallest $(\rho(E_F)\Delta^2(E_F))^{-1}$ values are obtained for Pauli-like compounds such as  CeRh$_2$ and CeNi$_2$.  Systems which seem not to follow the trend are CeSb and CePd$_3$ (not shown in the figure); cf. Fig.\,\ref{fig:TK} and Tab\,\ref{tab:kondo}. Both systems are barely metallic and
have a very small DOS at the Fermi level, and  together with
small hybridization at the Fermi level; $(\rho(E_F)\Delta^2(E_F))^{-1}$ basically diverges.  In case of CeSb the vanishing DOS($E_{\rm F}$) agrees with findings from literature and has been  obtained from  a DFT+U study, see Ref.\,\onlinecite{Price:00}.  CePd$_3$ has a pseudo gap at the Fermi level with intrinsic gap states and can be viewed as a Kondo insulator.\cite{Bucher:96}  The other extreme case which seems not to fit in the picture is  CePb$_3$ for which a Kondo temperature of 0.002 K has been obtained from magnetic susceptibility measurements fitted to the Lethuillier model.\,\cite{Lethuillier:78} This point has also not been considered in the linear fits. Summarizing this section, 
we use a large compound data analysis to address the correlations between the Kondo energy scale and the hybridization  function in conjunction with the corresponding DOS. We find trends that are
in agreement with the expected Kondo energy estimates,
exhibiting the expected exponential behavior.  
Fig.\,\ref{fig:TK} can serve as a guideline for deciding whether a material
qualifies as a Kondo system, or at least exhibits "normal" Kondo behavior, since such materials are expected to lie on the straight line of Fig.\,\ref{fig:TK}. 
In addition, Fig.\,\ref{fig:TK}
provides an easy estimate of the Kondo temperature of any compound, simply by evaluating of the hybridization function and DOS($E_{\rm F}$). 
We stress here that the confirmation of the trend comes from analysis of the entire data set, not from one single compound or some small selection of compounds.  Collective trends are revealed by looking at the statistics of all the compounds we have at hand.*

\section{Summary and Conclusion}
We have performed data mining on known (mostly cubic binary) Ce compounds and analyzed the degree of hybridization between the Ce 4f electron and the valence electrons of the system. 
We used the recently developed f-electron database \cite{IMS} to search for systematic correlations between electronic structure properties like hybridization and electronic radii and the observable properties such as volume and Kondo energy scale.  To our knowledge this is the first effort in this field addressing a search for systematic correlations in f-electron materials. 
Our goal has been to develop criteria from which to predict materials properties such as the itinerant or localized character of 4f compounds, and their Kondo temperatures. We were able to define systematic trends and correlations that are natural yet can only be confirmed by looking at the entire data set of Ce compounds. 

Since the strength of hybridization between the 4f and valence electrons of the non-Ce constituent defines, in large part the characteristic energy scales and as a consequence much of the exotic properties of Ce based compounds, e.g. heavy Fermion behavior, Kondo physics, and RKKY driven anti-ferromagnetic magnetism
the hybridization function can be viewed as a tool with which to classify these compounds. 
The localized/itinerant character of known systems is clearly reflected in the calculated hybridization energies, not only for cubic systems but also for materials with different lattice structure which have been included as test systems. Thus, the hybridization function could be an effective guide in the search for new materials with desired properties, being a direct link to the desired behavior and more quantifiable than other measures, such as the partial density of states or fat-band representation of the electron dispersion.
In the developing field of computational materials design, such an easy classification method
would be very welcome. To make use of this tool, we have to link the hybridization to a second quantity, preferably some observable,  which can be tailored by composition or strain if grown on a surface. Here, the volume was a natural choice because of the well-known volume dependence of the Ce mono-pnictides. Indeed, our results show systematic and significant anti-correlation between the hybridization and the volume or the related radius per atom  and a medium to high predictive power.  Interestingly,
in addition to the inverse relation between volume and hybridization,
our results show a fine-structure depending on the parity of the non-Ce valence electrons. The absolute values of the hybridization tend to be larger for systems with mostly p valence electrons (such as CeO$_2$ or CeSe) whereas for systems with s and d valence electrons such as CeM$_2$ (M = 3d transition metal) the hybridization is smaller.

Since the hybridization energy is an essential part of the Kondo coupling constant,
it is expected that the trends of the Kondo behavior should be reflected in the calculated data. Indeed we were  able to show that  the calculated  hybridization combined to the DOS agrees with the experimental findings for the Kondo temperature following the relation $\ln{T_{\rm K}}\,\sim\, 1/(\Delta^2(E_{\rm F})\,\rho(E_{\rm F}))$.  From the systems we have investigated so far, it seems clear that the outliers, \ie systems which do not show the linear dependence, are not really metallic Kondo compounds and should be characterized differently.

The three observations -- the anti-correlation, the valence electron dependent fine-structure, and the trends found for the Kondo temperature -- can  be used to find materials with desired properties.  
The main purpose of our approach is to find significant trends that can become apparent only by looking at the entirety of data. 
Although we focus here on Ce compounds, the approach suggested here can be used for other correlated electron systems, such as U and Pu based compounds, that potentially have an even more complex and intricate competition between energy scales, that  result in a plethora of exotic electronic states.

\section*{Acknowledgements}
The authors would like to thank Joe Thompson for fruitful discussions. 
O.E. and H.C.H. would like to thank Anna Delin for helpful discussions on the Kondo physics of the systems, and the values of 4f binding energies.
We acknowledge financial support from the Swedish Research Council. 
O.E acknowledges support from KAW (projects 2013.0020 and 2012.0031) as well as eSSENCE. H.C.H. acknowledges financial
support via STandUP. Work of AVB and TA was supported by UD DOE E3B7, KAW 2013.0096 and Villum foundation. 
The calculations were performed at NSC (Link\"oping University, Sweden), PDC (KTH, Stockholm, Sweden), HPC2N (Ume\r{a} University, Sweden) under
a SNAC project and at LANL facilities.

\end{document}